# The increased efficiency of fish swimming in a school

C.K. Hemelrijk[1*], D.A.P. Reid[1], H. Hildenbrandt[1] and J.T. Padding[2]

[1] Behavioural Ecology & Self-organization, Rijksuniversiteit Groningen, Nijenborgh 7, Groningen, The Netherlands

[2] Multiscale Modeling of Multiphase Flows, Eindhoven University of Technology, Eindhoven, The Netherlands

* Corresponding author: c.k.hemelrijk@rug.nl

**Summary**

There is increasing evidence that fish gain energetic benefits when they swim in a school. The most recent indications of such benefits are a lower tail (or fin) beat at the back of a school and reduced oxygen consumption in schooling fish versus solitary ones. How such advantages may arise is poorly understood. Current hydrodynamic theories concern either fish swimming side by side or in a diamond configuration and they largely ignore effects of viscosity and interactions among wakes and individuals. In reality, however, hydrodynamic effects are complex and fish swim in many configurations. Since these hydrodynamic effects are difficult to study empirically, we investigate them in a computer model by incorporating viscosity and interactions among wakes and with individuals. We compare swimming efficiency of mullets of 12.6 cm travelling solitarily and in schools of four different configurations at several inter-individual distances. The resulting Reynolds number (based on fish length) is approximately 1150. We show that these fish always swim more efficiently in a school than alone (except in a dense phalanx). We indicate how this efficiency may emerge from several kinds of interactions among wakes and individuals. Since individuals in our simulations are not even intending to exploit the wake, gains in efficiency are obtained more easily than previously thought.

## Introduction

There is an increasing number of studies that show energetic benefits for fish swimming in a school versus solitarily (1-6). A theory is needed that indicates whether energetic benefits are specific to certain configurations and which ones, or whether energetic benefits are a general outcome of proximity between the fish. Such a theory is important not only scientifically, but also technologically because it may have implications for the optimisation of coordinated swarms of robots that are used for surveillance of the environment underwater (7)

Empirical studies show that travelling in a school benefits individuals energetically, as demonstrated by a reduction of oxygen consumption in the school and of the frequency of beating a tail and fins by individuals at the back of the school (1-6). The mechanics underlying energetic benefits in a school may be related to the observation of reduced muscle activity when fish are slaloming between the vortices in the von Kàrmàn street or drag wake behind a pillar (8). This has also been suggested in an abstract, mathematical model by Weihs of a two-dimensional school that is constrained to a single plane. he argues that it is beneficial for a third fish when it swims midway between two preceding fish that swim next to each other, because at this location a von Kàrmàn vortex street or drag wake is also supposed to arise (9). In an infinite school, therefore, a diamond-like configuration of the positions of individuals is predicted to be energetically optimal. In this configuration, individuals are also expected to profit from their lateral neighbors through a channeling effect or wall effect (9-11). Despite the elegance of this theory, there is little evidence that the predicted diamond-like configuration occurs in nature (12,13) and it is known that the spatial configurations of individuals in schools are manifold and often irregular. Because in Weihs' theory the fluid dynamics are strongly simplified, his theory may have been too abstract for its aim. Therefore, in the present paper we pose the question whether including in a model more of the complexity of hydrodynamic interactions in a school may result in different predictions of the energetic advantages of travelling in a school. Therefore, we perform a more detailed simulation that represents particulars of the fluid hydrodynamics, undulation and fish shape. Here, we compare the energetic efficiency of fish when swimming solitarily versus in schools of several spatial configurations.

To represent the fluid we use a mesoscale particle-based simulation model, called Multi-Particle Collision Dynamics (MPCD). This has several reasons. First, it realistically represents hydrodynamics. Second, its particle-based grid-free nature makes it particularly suited to the simulation of multiple organisms that are moving and deforming (14-17). Third, in our earlier work on a fish travelling solitarily, we have already demonstrated its accuracy in reproducing the fluid dynamics in terms of the size of vortices, their circulation and their frequency of shedding behind a solitary mullet by comparing these to experimental results, see Table IV in our earlier paper (18). Fourth, it has shown to be useful in generating new explanations that can be used as hypotheses to guide empirical studies (18). As to its accuracy, our previous MPCD model accurately predicts the lift and drag forces on several static shapes (19), and it reproduces the positive correlation between swimming speed $U$ of the fish and its tailbeat frequency (18,20). The model reproduces quantitatively the reverse von Kàrmàn street behind a mullet-like fish (18,21). Moreover, it shows several relationships regarding slip ratio and Strouhal number that we recently confirmed in a large scale meta-analysis of empirical data (18,22), namely the slip ratio $U/V$ increases with the swimming speed (where $U$ is the speed of the fish and $V$ is the speed of its body wave) and the Strouhal number St decreases with tail beat frequency (St=$2Af/U$ with $A$ representing the tail beat amplitude and $f$ the tail beat frequency), in line with results of others (18,22-25). As to its usefulness in generating new testable hypotheses, our model explains the increase in slip ratio with tail beat frequency as a consequence of the lower variation of instantaneous swimming speed, which may be due to the accompanying smaller deceleration during the reversal of the tail beat, which we attribute to the lower effect of viscosity at a higher Reynolds number (18,22). The decrease in Strouhal number with tail beat frequency, we attribute to the stronger increase with tail beat frequency of the lateral power $P_s$ than of forward thrust $T$. This increases the width of the wake more than the longitudinal distance between the vortices, meaning that vortices are closer together at higher tail beat frequency (18).

As in the theoretical work of Weihs, we simulate infinite schools of fish confined to a single two-dimensional plane, with individuals swimming in phase (9). We compare the advantages of swimming solitarily versus in schools with individuals distributed in four configurations: swimming in a line behind each other, in a side-by-side phalanx, a rectangular configuration and a diamond-like configuration, and at several distances to each other.

## Materials and methods

*Species*

We base our simulations on the shape and swimming characteristics of a mullet of 12.6 cm (*Chelon labrosus*) for several reasons. First, details of the wake of such a mullet and the kinematics of its swimming have been reported (21) and have been shown to be similar to that in our model of a single mullet (18) regarding swimming speed and wake structure, namely the ring radius and angle of the vortices, their circulation, the standard deviation of their positioning in the longitudinal and lateral direction and the Strouhal number. Second, when a mullet swims steadily at cruising speed, it relies on the undulation of its body without use of its pectoral fins (20). Third, a mullet is considered an obligate schooler (20). Fourth, computational models show similar results for this type of (carangiform) swimmer when modeled in two and in three dimensions (18,19,23,26). Remarkably, results of two-dimensional models of swimming fish, not only of ourselves (18,19), but also of others, resemble those of real fish remarkably(27,28).Therefore, and because it is computationally cheaper, we develop our simulation in two dimensions.

*The fluid model*

Here we describe the main aspects of our basic model of undulating fish moving in a fluid of which the hydrodynamics is represented by multi-particle collision dynamics (MPCD). For more details see supplementary material and our earlier description (18). In MPCD a fluid is represented by ideal point particles. After propagating the particles for a fixed time interval according to $\mathbf{x}_i(t+\Delta t) = \mathbf{x}_i(t) + \mathbf{v}_i(t)\Delta t$, the system is partitioned into cells, and in each cell the velocities are changed according to $\mathbf{v}_i = \bar{\mathbf{v}} + \mathbf{\Omega} \cdot (\mathbf{v}_i - \bar{\mathbf{v}})$. Here $\bar{\mathbf{v}}$ is the mean velocity of the particles in the grid cell and $\mathbf{\Omega}$ is a stochastic rotation matrix that rotates the velocities by either $+\alpha$ or $-\alpha$ (where $\alpha$ is a fixed system parameter), with equal probability. It is the same for all particles within a cell. The rotation procedure can thus be viewed as a coarse-graining of particle collisions over space and time. The procedure conserves mass, momentum, and energy, and quantitatively yields the correct hydrodynamic (Navier-Stokes) behavior at least up till a Re number of 1520 (14-17).

An advantage of the MultiParticle Collision Dynamics method is that analytical expressions are available for several of its transport properties; see supplementary material (14).

In our MPCD simulation we have optimized the parameters for swimming at a medium Reynolds number (~1150) by making the viscosity as small as possible (Table 1). We confirmed that at all along the length of the fish, the thickness of the boundary layer is much larger than the scale of resolution (the collision cell size) of the model. To prevent compressibility effects, the maximum surface velocity of the fish (at the tip of the tail) is made always less than 20% of the speed of sound in the fluid.

Note that the flow speed in the model fluid emerges entirely from the undulatory motion of the fish.

*The fish model*

We scale the fish in terms of fractions of its body length *L*, so that *x*=0 at the front of the fish and *x*=1 at its rear and the spine of the straight fish has an *y*-value of 0 (Fig.. 1). Over time the lateral deviation from each point *x* on the spine of the fish from the central axis is given by

$$y(x,y) = \theta(x)\sin(k_L x - \omega t). \tag{1}$$

Here $\theta(x)$ represents the amplitude envelope, which varies nonlinearly along the fish body, $k_L = \dfrac{2\pi}{\lambda}$ is the wave number, which indicates the number of complete sine waves on the body for a wavelength λ, and $\omega = 2\pi f$ is the angular velocity for the tailbeat frequency *f*. For undulating mullets the amplitude of the wave is smallest behind the head and increases quadratically towards the tail:

$$\theta(x) = \theta_0 + \theta_1 x + \theta_2 x^2, \tag{2}$$

where the coefficients are given by $\theta_0 = 0.02$, $\theta_1 = 0.08$ and $\theta_2 = 0.16$ after empirical data (29).

The interactions between fluid and the undulating bodies are modeled through particle-wall collisions that lead to no-slip boundary conditions (18). As in our earlier model (18), we represent the fish as a deformable polygon consisting of line segments (Fig. 1), and parameterize the contours and (fixed) tail beat frequency after empirical data of Müller et al (21).

It is important to note that the swimming speed of the fish (both forwards and sideways) emerges entirely from the reaction force of the fluid caused by the prescribed undulatory motion of the fish.

*Different school configurations and distances among individuals*

To simulate an infinite school, we place one or two fish in a simulation area of width *W* and length $L_{\text{Area}}$ (Fig. 2a) and apply periodic boundary conditions. We organise the individuals in a diamond configuration, as studied by Weihs, by placing two individuals in diagonally opposite corners of the simulation area (Fig. 2a). We generate a rectangular configuration by placing a single fish in the centre of the simulation area (Fig. 2b). To obtain schools that are infinite in only one direction, we apply periodic boundaries in one direction and scramble the fluid along the boundaries of the other direction. Our methods of scrambling remove the structure and vorticity in the flow by resetting the positions and velocities of fluid particles to those of a homogeneous fluid. This is done by redistributing the current velocities of fluid particles randomly over new positions along the width of the wall. This procedure thus conserves the average velocity and maintains the average momentum flux across the boundary, as described earlier (18). Scrambling the boundaries ahead and behind a single fish produces a phalanx, whereas scrambling the boundaries lateral of the individual leads to a line configuration (Fig. 2c, d). We study the hydrodynamic consequences of different inter-individual distances in all four configurations. For the line configuration we study longitudinal distances ranging from 0.5 to 5 fish lengths *L*. For all other configurations, we keep the longitudinal distance fixed at the average distance for empirical data, namely one fish length *L* (1-6) and we study lateral distances ranging from 0.4 fish lengths *L* (the distance considered optimal by Weihs) to 2*L* which is the maximum possible on our hardware.

*Measurements*

There is debate about how to measure swimming efficiency (30). One frequently used measure is the Froude efficiency, a dimensionless number, indicating the quotient of the useful power divided by the total power. If, however, useful power is based on the net thrust (thrust plus drag) during steady swimming, which is zero, then Froude efficiency is zero. Although thrust cannot be measured for steady swimming empirically, it can and has been measured in models, such as Elongated Body Theory and numerical models, like ours, see figure 14 of our earlier paper (18). The Froude propulsive efficiency separates for a steady swimming fish the useful power that the fish exerts forward (thrust) from the wasted power it exerts sideward (due to its sideways tail movement), thus calculating the fraction of the total power spent by the fish which is converted into forwards speed (23,31). We use this modified version of the Froude efficiency $\eta$:

$$\eta = \frac{\overline{T}\,\overline{U}}{\overline{T}\,\overline{U} + \overline{P}_s}, \quad (3)$$

with $\overline{T}$ the average thrust, $\overline{U}$ the average forward speed (of the fish relative to the fluid) and $\overline{P}_s$ the average lateral power per individual fish all taken over the beat cycle. Note that we need to take the average over the tail beat because even in steady state the velocity of the fish oscillates during each tail beat around its forwards average $\overline{U}$.

We measure the thrust *T*, sideways power $P_s$, the forward speed of the fish *U* and Froude efficiency $\eta$ after running the model until its steady state. We calculate average data over 20 tail beats.

The most relevant velocity for the estimation of the drag and efficiency is clearly the relative velocity between the fish and the water from the perspective of the fish. We name this speed *U* $U = U_{flow} - U_{fish}$. Here, $U_{flow}$ and $U_{fish}$ are both measured in the global frame, whereby $U_{flow}$ is the speed of the oncoming flow ahead of it (averaged over a sufficiently large area in front of the fish for its estimation, as indicated in Fig. 1, and $U_{fish}$ is the speed of the fish.

To measure thrust and sideways power, we decompose the forces on the skin of the fish. The total forward force *F* at any moment *t* is calculated by a summation over the line segments of the skin of the fish (Fig. 1, 3):

$$F(t) = \sum_i \mathbf{F}_i \cdot \mathbf{e}_f = \sum_i \left( \mathbf{F}_n^i \cdot \mathbf{e}_f + \mathbf{F}_t^i \cdot \mathbf{e}_f \right), \quad (4)$$

where $\mathbf{F}_n^i = (\mathbf{F}_i \cdot \mathbf{n})\mathbf{n}$ is the force vector perpendicular to the skin at line segment *i*, $\mathbf{F}_t^i = \mathbf{F}_i - \mathbf{F}_n^i$ is the force vector tangential to the skin and $\mathbf{e}_f$ is a forwards-pointing unit vector. In the summation in Eq. (4) the first element thus represents the contribution to the forward force of pressure and the second element that of the viscous force.

To separate the thrust and drag out of this total body force we decompose it, depending on whether or not the forward force is positive (thrust) or negative (drag) (23):

$$T(t) = \sum_i \left[ \mathbf{F}_n^i \cdot \mathbf{e}_f H(\mathbf{F}_n^i \cdot \mathbf{e}_f) + \mathbf{F}_t^i \cdot \mathbf{e}_f H(\mathbf{F}_t^i \cdot \mathbf{e}_f) \right]$$
$$D(t) = -\sum_i \left[ \mathbf{F}_n^i \cdot \mathbf{e}_f H(-\mathbf{F}_n^i \cdot \mathbf{e}_f) + \mathbf{F}_t^i \cdot \mathbf{e}_f H(-\mathbf{F}_t^i \cdot \mathbf{e}_f) \right]$$
(5)

where $H$ is the Heaviside step function. For each edge, for both the perpendicular (pressure) and tangential (viscous) force on it, we add the forward component of the force to the thrust if it is positive, and to the drag if it is negative. The sum of thrust and drag is the total force $F(t)$ along the $\mathbf{e}_f$ direction:

$$F(t) = T(t) - D(t),\qquad(6)$$

i.e. a positive $F(t)$ indicates net acceleration and a negative $F(t)$ a net deceleration in the $\mathbf{e}_f$ direction. In steady state, the time average of $F(t)$ over a full swimming cycle is zero.

We calculate per time step $\Delta t$ the sideways power $P_s$ that the fish exerts:

$$P_s(t) = \sum_i \mathbf{F}_i \cdot \mathbf{e}_s V_{und}^i,\qquad(7)$$

where $\mathbf{e}_s$ is a unit vector in the sideways direction, and $V_{und}^i$ is the sideways velocity of edge $i$.

## Results

Our model shows that for all configurations of schooling (apart from the phalanx) the Froude efficiency of individual fish (which is the percentage of total power converted in forward speed) is higher than for a solitary individual (Fig. 4a) and so is the speed (Fig. 4b), while thrust and sideways power are lower in all configurations in a group (apart from the phalanx) versus when swimming solitarily (Fig. 4c, d). We explain these findings by starting from the simplest configurations, the line and the phalanx.

The higher efficiency $\eta$ in the line-shaped school compared to a solitary fish is surprising, because when fish are swimming behind each other they are expected to slow down since they swim in the thrust wake of the predecessor and receive the jet of the predecessor on their nose . Our model shows that due to the undulation both the head of the follower and the jet of the predecessor move laterally back and forth whereby the jet passes the body sometimes to the right, sometimes to the left and the individual sometimes captures vortices in a beneficial way (Fig. 5, 6a, Movie S2). When the fluid passes the body at one side, it herewith causes a pressure difference, so that the individual experience lift to the side where the fluid passes faster. Consequently the individual will experience a lateral force and indeed there is an overall net drift laterally whereby the direction depends on the direction of the first beat (Fig. S1 Suppl mat).

The Froude efficiency of fish swimming in a (not too dense) phalanx is higher than when swimming alone which is in line with the channeling effect (Fig. 4a). Surprisingly, the Froude efficiency of the phalanx is lower than that of the line (Fig. 4a, Movie S3). This is probably due to an increased resistance (on average per fish) of the phalanx to oncoming flow due to close proximity of lateral neighbours (for all lateral and longitudinal distances measured here). In agreement with this, the speed of the phalanx is usually lower than that of a solitary fish (Fig. 4b) and at larger lateral distances among individuals (where presumably resistance to oncoming flow is lower), both efficiency and speed increase not only in a phalanx, but also in a rectangular and diamond-shaped configuration (Fig. 4ab).

The rectangular configuration is more efficient than both the line and the phalanx. This may arise because in the rectangular configuration individuals benefit from neighbours in both directions, the longitudinal and the lateral one (Fig. 4a, 5, Movie S4).

Remarkably, at short lateral distances (at 0.4 *L*) the diamond-shaped configuration is less efficient than the rectangular configuration and fish even swim slower than a solitary fish (Fig. 4ab). This may arise because the fish receives an almost laminar structure in front of it (black rectangle Fig. 6b). This happens, because the undulating bodies of the neighbours one row (diagonally) ahead compress and disturb the wake of the predecessor straight ahead (two rows ahead, Fig. 6b, Movie S5). Consequently, individuals benefit longitudinally less than in the rectangular configuration where they receive the wake of their predecessor on their laterally moving head. At larger lateral distances in the diamond configuration, however, the fish encounters a more intact wake structure generated by the predecessor straight ahead (Fig. 5c). Here the "diamond effect" of swimming midway between others may work. This may explain why at large lateral distances, above 0.4 *L*, the efficiency of a diamond-like configuration is similar to that of a rectangular configuration (Fig. 4a).

## Discussion

Our model confirms and rejects a number of theories and generates new ones.

It confirms the higher efficiency of swimming in a diamond configuration and a phalanx (apart from the densest phalanx) compared to swimming alone (9)

Unexpectedly, in contrast with expectations based on the rotations of vortices in a thrust like vortex system (9), it is efficient to swim directly behind another fish. Our model shows that this arises because following fish move their head sideways and therefore, they do not exactly swim in a thrust wake (experiencing only disadvantageous flow from the vortices). Instead they may sometimes, by bending their head sideways when a vortex arrives, capture it such that they benefit from its flow. We may speculate that real fish may even exaggerate their lateral movements in a way synchronized to the shedded vortices to capture their energy, thus performing a kind of reverse von Karman gait (32) .. Note that the beneficial effects decrease with distance as expected, but that even at the largest distance studied here they do not yet reach the values of the single fish, due to the slow decay of hydrodynamic flow. In a dense diamond-like configuration, however, close lateral proximity among neighbours ahead distorts the otherwise incoming vortex wake, and ruins its beneficial effect by producing an almost laminar structure. Therefore, our model rejects that the optimal lateral distance of the diamond configuration should be 0.4 *L* as predicted by Weihs (9). The difference between our results from Weihs' theory (9) is due to the different representation of hydrodynamics in both models; our present model takes into account viscosity, deformable body shape and interactions between wakes and individuals, none of which the theory of Weihs represents.

New is further that our model shows that swimming in a school is almost always more efficient than swimming solitarily, irrespective of the precise spatial configuration of the school (also when fish swim in a line and rectangular configuration) and almost independent of the lateral and longitudinal distance among individuals in our study, and apparently irrespective of the phase of the wake when it is received. Thus, in a school the interaction between water and fish enhance forward movement in more ways than thought before, not only by exploiting a reverse von

Karman street or drag wake and by the canalization effect, but also by exploiting through the sidewards movement of the head a reverse von Karman street or thrust wake. Our results raise many further questions. How will the beneficial effect of schooling be influenced when, first, the size of the school is made finite, second, when individuals undulate asynchronously, third, when individuals actively exploit flow by adjusting their position and undulation, and fourth, when individuals actively coordinate with others (33,34)? New simulations are planned to study these questions.

In sum, incorporating in a computational model bodily movements, viscosity and interactions among wakes and individuals leads to novel findings regarding hydrodynamic advantages of schooling. Our model suggests that schooling is in general beneficial. Yet in order to know whether this holds in reality also, we still need to study the effects of the natural variation among individuals in their distances and angles to each other, their body sizes, their swimming phases and their tail (or fin) beat (1-6). Since benefits in efficiency in our simulations arise while individuals are not even intending to exploit the wake, hydrodynamic benefits are easier to obtain than previously thought and are therefore an important topic for further investigation.

## Acknowledgments

We are grateful to the positive receipt of our work by Prof. dr. D. Weihs. Our studies have been inspired by the enthusiasm about hydrodynamics of Prof. dr. E. J. Stamhuis and Prof. dr. J. J. Videler.

## Financial support

We thank the Groningen University for an Ubbo Emmius grant to Daan Reid, and the Dutch Organization for Scientific Research (Nederlandse Organisatie Wetenschappelijk Onderzoek, NWO) NWO *Cognitie Pilot* project (051.07.006) for financial support to Hanno Hildenbrandt.

*Tables*

| Property | Sim. Value | Value |
|---|---|---|
| Particles per cell (avg.) $\gamma$ | 8 | - |
| Rotation angle $\alpha$ | $\pi/2$ | - |
| Viscosity of fluid $\mu$ | 1.15 $m/(a_0\Delta t)$ | $10^{-3}$ kg/(m.s) |
| Fish length $L$ | 900 $a_0$ | 0.126 m |
| Number of edges per fish $N_E$ | 1024 | - |
| Simulation box length $L_{area}$ | ??? $a_0$ | ??? m |
| Wave number fish deformation $k_L$ | 1.8 $\pi/L$ | 1.8 $\pi/L$ |
| Tailbeat frequency $f$ | $5.7 \times 10^{-4} / \Delta t$ | 3.8 $s^{-1}$ |
| Tailbeat amplitude $A$ | 90 $a_0$ | 0.0126 m |
| Reynolds number | 1150 | 1150 |

Table 1. Parameters of the fluid and fish in the Multi-Particle Collision Dynamics simulation. For units of length, mass and time we choose the collision cell size $a_0$, particle mass $m$ and collision time step $\Delta t$, respectively (second column). Real (SI) values (last column) are obtained by substituting $a_0 = 1.4 \times 10^{-4}$ m, $m = 1.83 \times 10^{-11}$ kg, and $\Delta t = 1.5 \times 10^{-4}$ s.

*Figures*

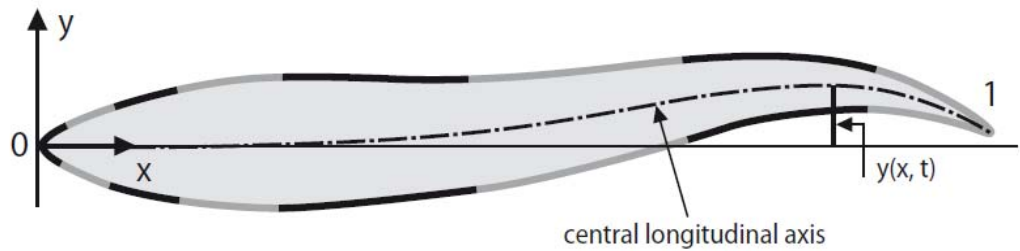

Figure 1. Schematic overview of the deviation from the central axis of the spine of an undulating mullet and of the representation of the mullet as a deformable polygon comprising line segments (indicated in grey and black). Note that line segments on the head are shorter than elsewhere on the body.

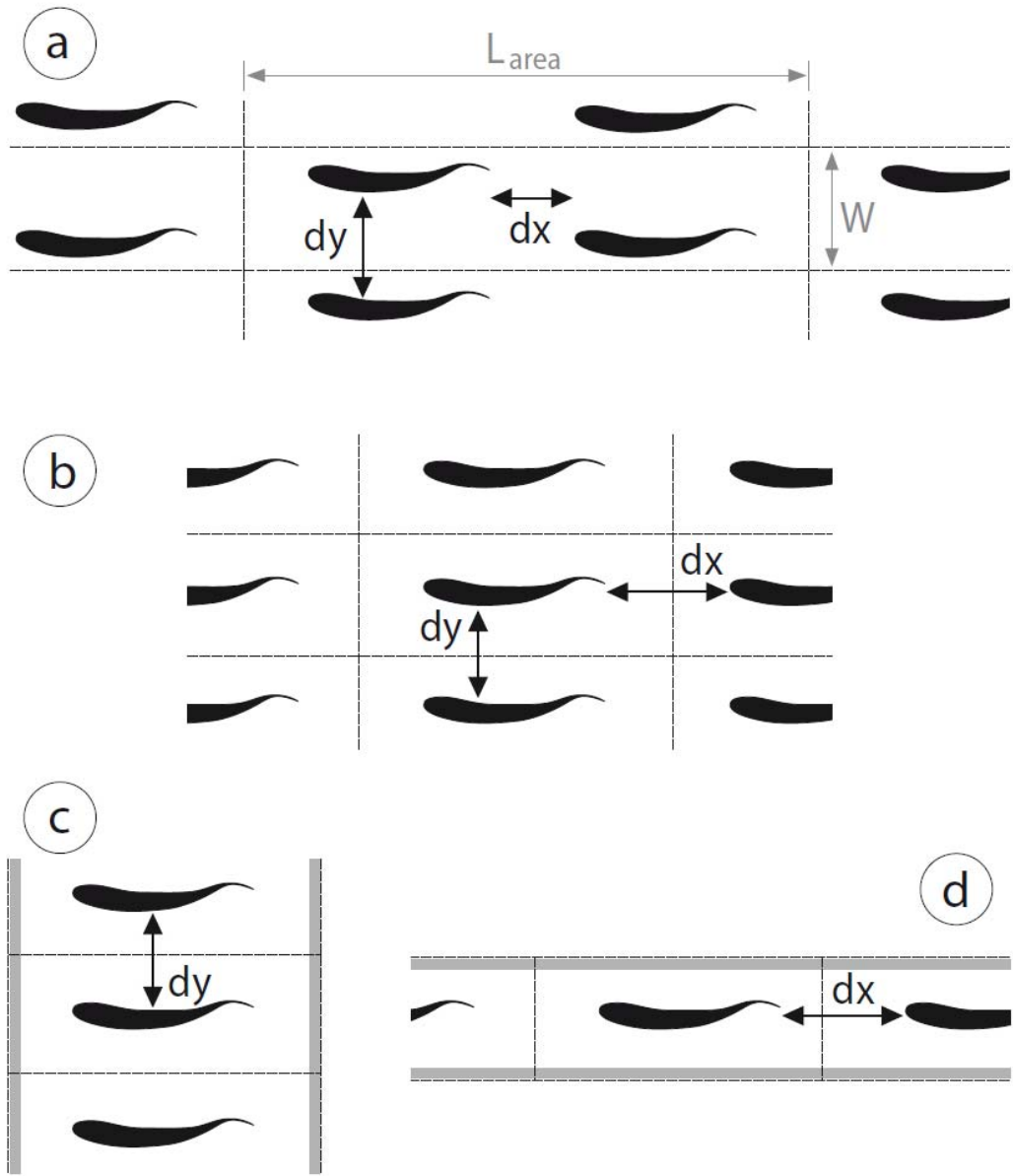

Figure 2. Periodic 2-dimensional and 1-dimensional arrangements of fish. a) diamond, b) rectangular, c) phalanx and d) line configuration. The discontinuous outline indicates the simulation area. Scrambled boundaries are marked by a wide band of light gray. $L_{area}$ indicates the length of the simulation area, W indicates its width. Lateral distance between neighbours is given by *dy* and longitudinal distance by *dx*.

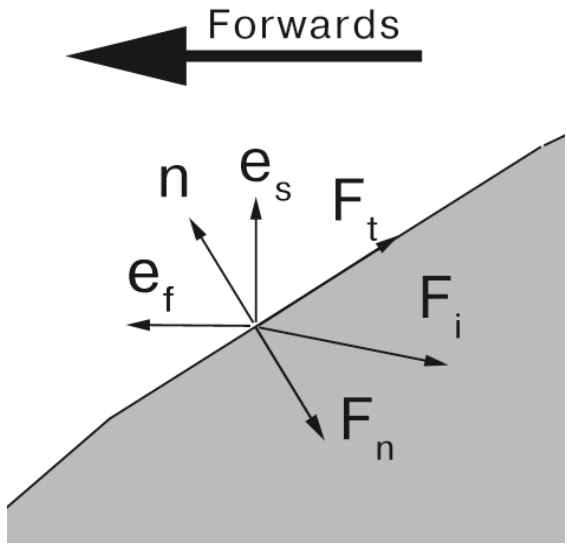

Figure 3. Decomposition of the force $F_i$ on line segment $i$ of the skin of the fish into pressure ($F_n$) and viscous ($F_t$) components. The surface normal is indicated as and the unit vectors pointing forward and sideways are labeled $e_f$ and $e_s$.

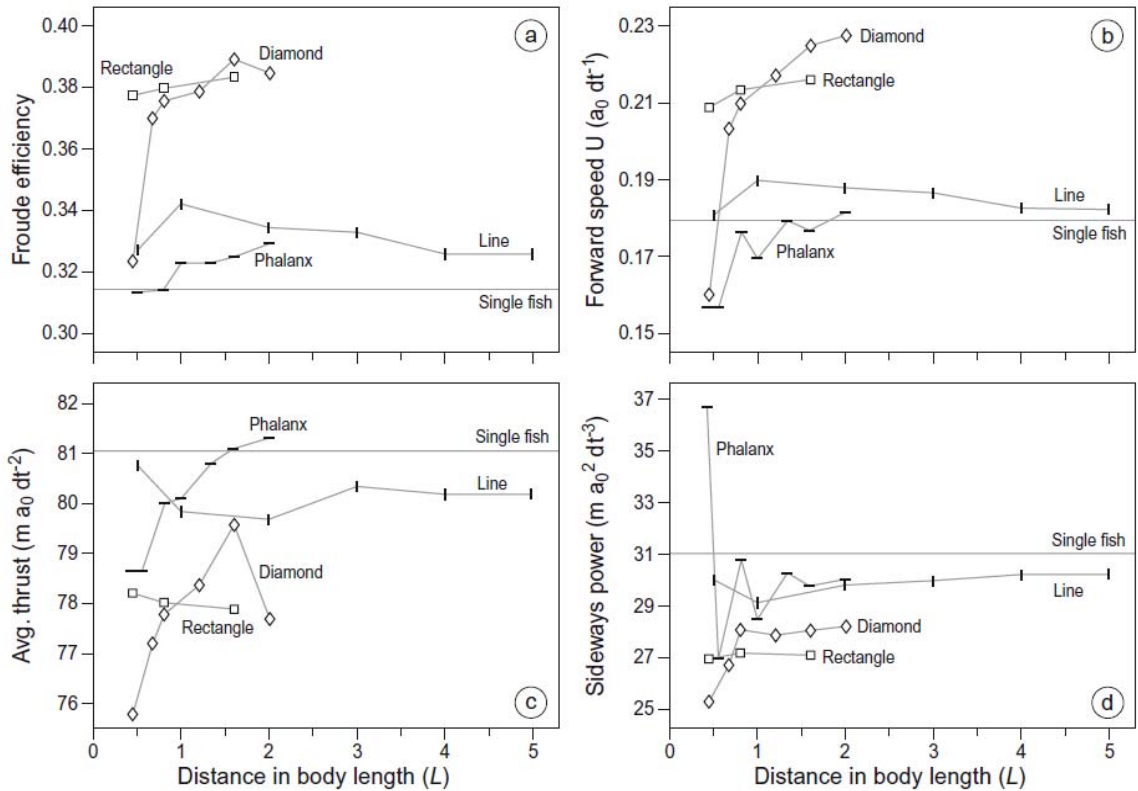

Figure 4. Froude efficiency $\eta$ (a), speed (b), average thrust (c) and sideways power (d) of different configurations for different distances (in body length $L$) among individuals. The phalanx, rectangular and diamond configuration are studied at several lateral distances ($dy$) and the line formation at several longitudinal distances ($dx$). Note that due to the large sample size (20 tail beats or 130.000 time steps) the standard error is essentially zero.

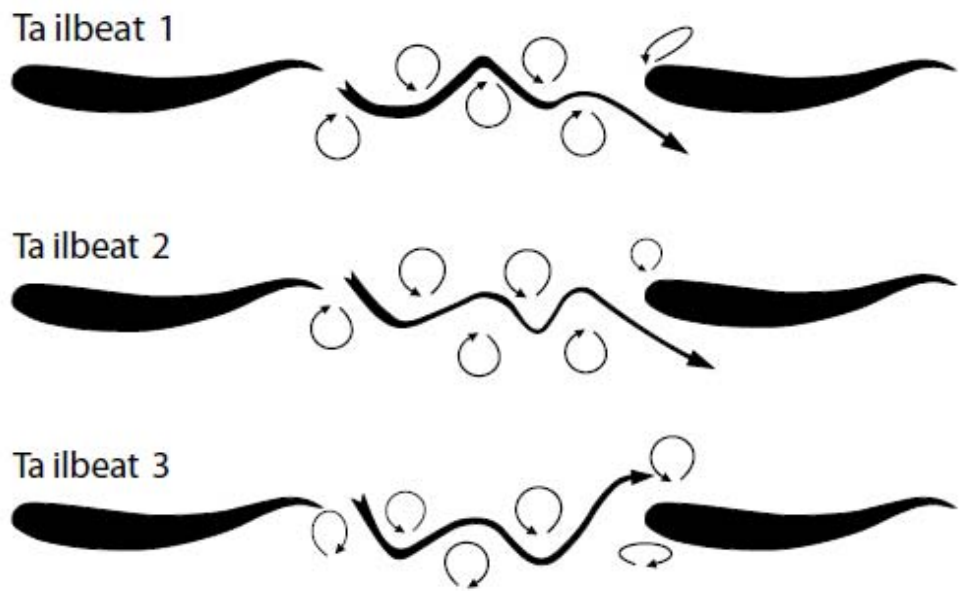

Figure 5. The structure of the wake in the rectangular configuration over three consecutive tail beats. Arrows give the direction of local flow and their thickness indicates the speed of local low. The fat, long arrow represents the jet. Information for the arrows was taken from snapshots of the flow field as in Fig. 6.

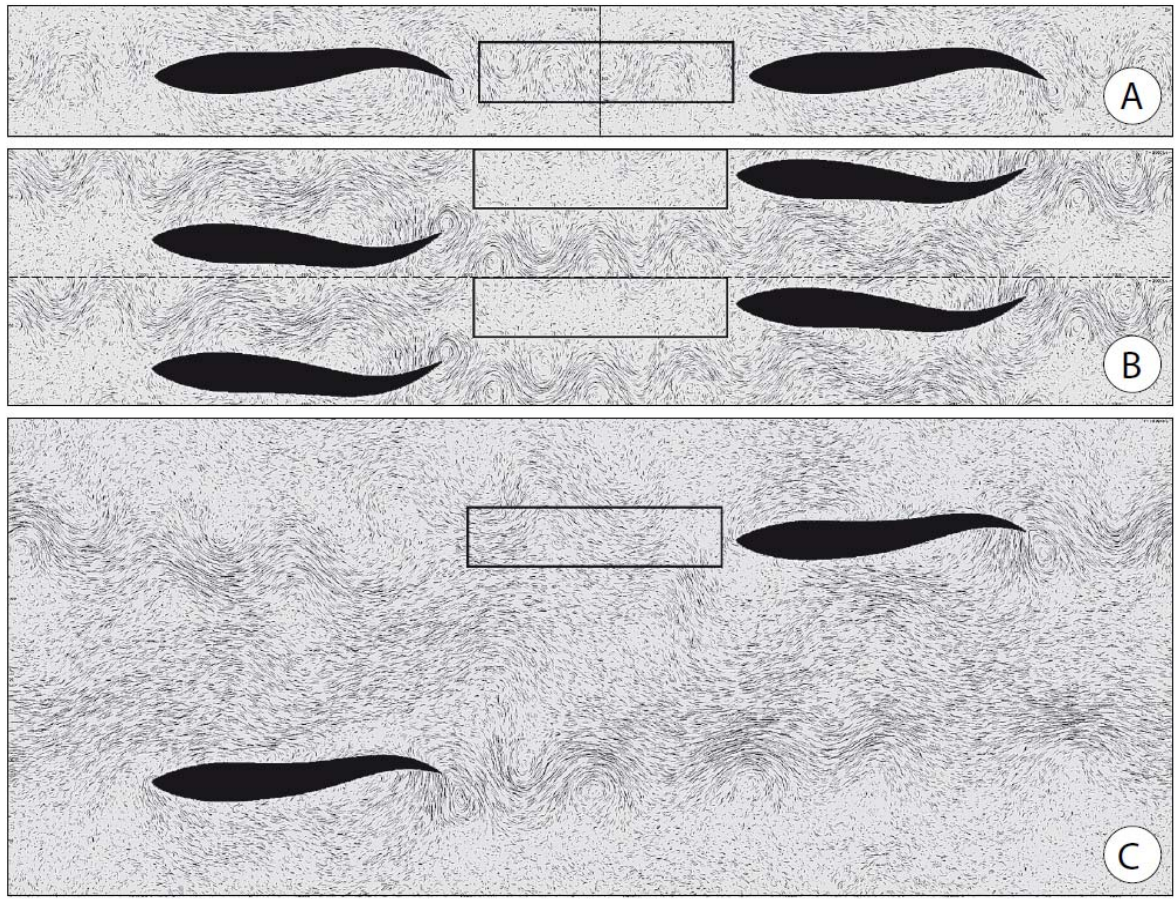

Figure 6. Flow in an infinite periodic school for several configurations. a) rectangular configuration with small lateral distance (0.4 $L$), b) diamond configuration with small lateral distance (0.4 $L$), and c) diamond configuration with large lateral distance (1.6 $L$). For clarity the simulation area is shown twice in Figs. a and b, with horizontal duplication in a and vertical in b as indicated by the dotted line. Lines represent truncated streamlines, longer lines indicating faster speed. Areas ahead of each fish are marked by a black rectangle. For videos of moving fish, see the supplementary material.